# High-efficiency 2D grating design for the magneto-optical trap: enhancing intensity balance and reducing optical complexity

ELAHEH KAROOBY[1], JIAZHEN LI[1], AMIT AGRAWAL[2], QING GU[1,3,*]

[1]Department of Electrical and Computer Engineering, North Carolina State University, Raleigh, North Carolina 27695, USA

[2]Electrical Engineering Division, Department of Engineering, University of Cambridge, Cambridge, CB3 0FA, UK

[3]Department of Physics, North Carolina State University, Raleigh, North Carolina 27695, USA*

*qgu3@ncsu.edu*

**Diffraction gratings integrated into an atomic, molecular, and optical (AMO) setup offer a compact and efficient route toward atom cooling and trapping, thus preparing magneto-optical traps (MOT) for insertion into future scalable quantum systems. Here, we propose and numerically validate a two-dimensional (2D) diffraction grating that satisfies the required optical conditions for laser cooling, namely, radiation pressure balance, specular reflection cancellation, and circular polarization handedness reversal upon diffraction, thus achieving an optical molasses – a necessary condition in MOT. Using Rigorous Coupled Wave Analysis (RCWA) and a Genetic Algorithm (GA), we optimize the grating's geometry to maximize key figures of merit. The grating consists of a 2D square lattice of nanoholes and is designed to diffract normally incident 780 nm circularly polarized light into the four first orders, each with a diffraction efficiency of 0.24, of which 99.7% has the correct circular handedness. Our 2D diffraction grating can enhance the performance of GMOTs employing 2D gratings by improving the axial radiation pressure balance and providing a high degree of circular polarization. Furthermore, we investigated the robustness of our 2D grating to ensure high manufacturing resilience.**

Magneto-optical traps (MOTs) are widely used to trap and cool neutral atoms to sub-millikelvin temperatures using a combination of laser beams and magnetic fields [1]. A MOT is a key tool for preparing ultracold atomic gases, which are crucial for exploring fundamental physics and technologies such as atomic clocks [2], quantum computing and simulation [3], and atom interferometry [4]. A MOT typically uses six or four circularly polarized laser beams with appropriate handedness directed toward the trapping region. These lasers are red-detuned from the targeted atomic transition by a few natural linewidths (Γ). As a result, an atom experiences Doppler cooling as it moves toward each beam. Simultaneously, a spatial gradient magnetic field, created using a pair of anti-Helmholtz coils, provides a Zeeman shift to the atoms, pushing the atoms toward the center of the MOT. Together, a trapping region is formed near the zero gradient quadrupole magnetic field in the overlap region of multiple circularly polarized, red-detuned beams.

With the advent of nanofabrication, significant efforts have been made to miniaturize MOTs using microfabricated atom chip traps. Two main types of miniature MOTs are pyramid MOT (PMOT) [5] and grating MOT (GMOT) [6–8]. A PMOT utilizes a single circularly polarized input beam to form the optical trapping region, simplifying optical alignment by requiring only one laser beam. However, PMOTs are limited by the low atom capture capability (<7 × $10^3$ atoms) [9]. Therefore, PMOTs are not suitable for applications requiring high signal-to-noise ratios and atomic degeneracy, such as quantum information processing [10], and high-resolution spectroscopy [11], which require >$10^7$ atoms. Although GMOTs operate on a similar principle to PMOTs, they overcome the low atom capture limitation by providing a larger overlap region, thanks to multiple diffraction orders from a large-area grating [12]. This enables GMOTs to trap $10^7$ - $10^8$ atoms, four orders of magnitude higher than PMOTs [12,13].

Two types of planar grating chips have been employed in GMOTs: segmented tri-gratings that consist of three one-dimensional (1D) gratings that are $2\pi/3$ angle rotated with respect to each other and extended towards the center of the chip [14–16], and 2D gratings [17–19]. For both grating geometries, chips with a 2 × 2 $cm^2$ area illuminated by a 2 cm diameter laser beam can provide optical overlap volumes of approximately 1 $cm^3$, significantly larger than the 0.01 $cm^3$ overlap volume of PMOTs [12]. However, the segmented tri-grating offers a smaller optical overlap volume than a 2D grating and, therefore, traps about one-third of the number of atoms [12]. In addition, there is a point of symmetry in the segmented tri-gratings' overlap volume, requiring precise alignment from the incident beam to the trapping point [20]. In contrast, the 2D grating has a larger overlap volume and does not require precise alignment, as the grating profile is uniform across the entire chip surface. However, 2D gratings' low diffraction efficiency prevents them from meeting the required balanced optical intensity along the vertical axis, necessitating the use of a neutral density (ND) filter to achieve this balance [17]. Therefore, most state-of-the-art GMOTs use 1D gratings instead of a 2D grating to avoid optical complexity. Ref [21]optimized and fabricated a four-segment grating for GMOTs to set the targeted first-order diffraction efficiency to 25%, and ref [22] proposed a grating design methodology and utilized it for the optimization of segmented tri-gratings for laser cooling.

It is therefore important to design high-efficiency 2D gratings that simultaneously satisfy the required optical conditions, allow easy alignment, and can trap a large number of atoms. Previous studies have clearly identified the necessary optical conditions for efficient Doppler cooling and trapping, including optical intensity (radiation pressure) balance and the circular polarization handedness reversal upon diffraction [5,6,8].

In this work, we design and optimize a 2D dielectric diffraction grating with a metal back reflector that operates at 780 nm for $^{87}$Rb GMOTs. The optimized 2D grating can simultaneously support high first-order diffraction efficiencies, zeroth order (specular reflection) cancellation, and high fidelity in maintaining the correct circular polarization handedness upon diffraction. Under normal incidence, the grating achieves a diffraction efficiency of 0.24 for each of the $(\pm 1, 0)$ and $(0, \pm 1)$ diffraction orders, with 99.7% of the diffracted beams with the correct circular handedness. This design can enhance the performance of GMOTs employing 2D gratings by improving the axial radiation pressure balance and providing a high degree of circular polarization. Furthermore, we examine the gratings' manufacturing tolerance, demonstrating the robustness of our design to fabrication imperfections. We believe this work significantly improves over previously reported 2D gratings for GMOTs, paving the way for next-generation chip-scale MOTs requiring high atom numbers.

Fig. 1(a) illustrates the GMOT configuration, where a 2D grating diffracts normally incident circularly polarized beam into four beams of first-order diffraction (Fig. 1(b)), forming a five-beam arrangement. The atom cloud can be trapped in the overlap region between the incident beam and the diffracted beams, positioned near the zero of the quadrupole magnetic field. When a laser beam is normally incident on the 2D grating, the diffraction angle $\theta$ – measured from the normal of the grating surface – for each order ($m$, $n$) is determined by the grating period $\Lambda$ (considering $\Lambda_x = \Lambda_y = \Lambda$) relative to the incident wavelength $\lambda$, according to the Bragg condition, $\Lambda \sin(\theta) = \left(\sqrt{m^2 + n^2}\right)\lambda$. When $\Lambda$ is restricted to the range $\lambda < \Lambda < \sqrt{2}\lambda$, the optical power is primarily distributed among the (0, 0), (0, ±1), and (±1, 0) diffraction orders. For an incident wavelength of 780 nm, corresponding to the D2 transition ($5S_{1/2}$ to $5P_{3/2}$) of $^{87}$Rb, we design the grating period $\Lambda$ to be 1080 nm, resulting in a diffraction angle of $\theta$ = 46.24°, which has been experimentally shown to be appropriate for GMOTs [12]. Although the optimal diffraction angle $\theta$ for GMOTs cannot be analytically determined, it must meet a few specific requirements. First, $\theta$ must be greater than 30° to avoid high diffraction orders. Second, a small $\theta$ leads to low radial trapping forces and reduces cooling and trapping efficiency [5]. Conversely, a very large $\theta$ significantly limits the trapping volume $V$, formed by the intersection of the incident and diffracted beams, thereby reducing the number of trapped atoms, $N_{atoms} \propto V^{1.2}$ according to an approximate scaling law [23]). The volume of the trapping region formed by the diffracted beams of a 2D grating with side length $a$, illuminated by a laser beam of diameter $2a$, can be approximated as the volume of a square pyramid with height $h$ ($V = (1/3)a^2 h$), where $h$ is determined by $\theta$ [12].

Moreover, to achieve balanced optical molasses, the optical forces acting on the atoms should sum to zero, both radially and vertically, at the center of the quadrupole magnetic field. Thanks to the symmetry of the 2D grating, radial optical balance is ensured, and the largest overlap volume is achieved when the incident beam is aligned with the center of the grating. However, the balance along the vertical axis is non-trivial, because GMOTs spatially compress the incident beam upon diffraction, increasing the beam intensity in each diffraction order [6].

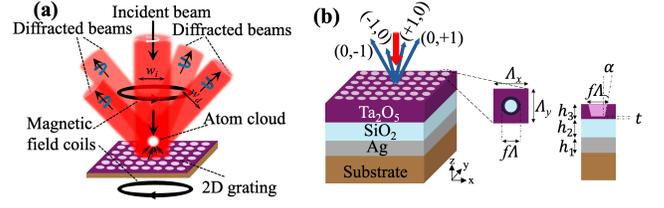

Fig. 1. A schematic of: (a) atom cooling and trapping using a 2D diffraction grating atom chip. Black arrows point out the propagation direction of the incident and four first-order diffracted beams, with circular polarizations indicated by white and blue arrows, respectively. (b) the proposed 2D diffraction grating structure, with a top view and side view of a unit cell.

For an incident beam of intensity $I_i$, the intensity of each first-order diffracted beam is $I_d = \eta_{(1,0)}(w_i/w_d)I_i = \eta_{(1,0)}(I_i/\cos\theta)$, where $\eta_{(1,0)}$ is the efficiency of each first-order diffracted beam, $w_i$ and $w_d$ are the beam waist of the incident and first-order diffracted beam, respectively, as illustrated in Fig. 1(a). For $N$ first-order diffracted beams, the total upward intensity is $NI_d \cos\theta = N\eta_{(1,0)}I_i$ and the condition for vertical intensity balance require that $\eta_{(1,0)} = (1/N)$. Considering the net intensity incident on the grating chip as $I_i(1 - \eta_{(0,0)})$, where $\eta_{(0,0)}$ represents the efficiency of the (0, 0) order, the vertical intensity balance can be quantified using a dimensionless parameter known as the radiation balance, $\eta_B = \frac{N\eta_{(1,0)}}{(1-\eta_{(0,0)})}$ [18], ideally equal to 1.

Another optical requirement of the grating concerns the polarization of the diffracted beams. To facilitate efficient atomic transitions, the first-order diffracted beams must exhibit circular polarization with handedness opposite to that of the incident circularly polarized beam, aligning with the Zeeman effect in the atom's interaction with the magnetic field. In contrast, the zeroth-order diffracted beam must maintain the same handedness as the incident beam to prevent anti-trapping. Expressing the electric field vector of each first-order diffracted beam with S and P polarization components of intensities $I_s$ and $I_P$ respectively, and a phase difference of $\varphi_{SP}$, we can write:

$$\begin{pmatrix} \sqrt{I_s} \\ \sqrt{I_p}e^{i\varphi_{SP}} \end{pmatrix} = \sqrt{I_+}e^{i\varphi_+}\begin{pmatrix} 1 \\ +i \end{pmatrix} + \sqrt{I_-}e^{i\varphi_-}\begin{pmatrix} 1 \\ -i \end{pmatrix} \quad (1)$$

Eqn. (1) can be re-written in terms of the contrast of oppositely handed circular polarizations, which either trap or anti-trap the atoms (denoted by + or – subscripts, respectively), representing the degree of circular polarization:

$$\xi_{SP} = \frac{I_+ - I_-}{I_+ + I_-} = \frac{2\sqrt{\frac{I_s}{I_p}}\sin\varphi_{SP}}{1+\frac{I_s}{I_p}}, \quad (2)$$

Ideally, $I_+ = 1$ and $I_- = 0$, resulting in $\xi_{SP} = 1$.

Illustrated in Fig. 1(b), the diffractive grating chip is based on a 2D square lattice of nanoholes within a tantalum pentoxide ($Ta_2O_5$) [24] layer of thickness $h_3$, which interfaces with a silicon dioxide ($SiO_2$) buffer layer of thickness $h_2$, mainly used for fine tuning the optical path length of the diffracted light, to achieve optimal reflection (ee Supplement 1). This buffer layer is deposited above a 200 nm silver (Ag) layer of thickness $h_1$, which functions as a highly reflective mirror around 780 nm. Zerodur is selected as the substrate due to its low thermal expansion coefficient and high resistance to deformation under various temperatures [25]. At 780 nm, the refractive indices of $Ta_2O_5$, $SiO_2$, and Ag are 2.1 [26], 1.45 [27], and (0.0905 + i5.0617)[28], respectively. For the top $Ta_2O_5$ nanohole 2D grating, we consider a realistic scenario of the etched nanohole exhibiting a tapered sidewall with tilt angle $\alpha$. To account for etching errors, we consider an unetched $Ta_2O_5$ layer of thickness $t$. We use the GD-Calc open-source code [29] based on RCWA to design the 2D grating (details in supplement 1).

Although the multilayer 2D grating geometry itself is not unique, this study is distinguished by its specific design requirements for GMOTs. 2D gratings have been designed for various applications, typically optimized to maximize a particular diffraction order under a specific incident polarization, usually linear.

For the 2D grating studied here, high-efficiency first-order diffracted beams are essential for improving optical intensity balance along the vertical axis. In contrast, the zeroth-order beam must be suppressed, as its circular polarization handedness is different from the incident beam, which can cause anti-trapping. Therefore, we need to optimize the grating structure to simultaneously maximize the first-order diffraction efficiency and minimize the zeroth-order efficiency, while ensuring that the first-order diffracted beams exhibit a high degree of circular polarization with the correct handedness. The geometrical parameters of the 2D grating, including the fill factor $f$ (the ratio of the hole diameter to the period), $SiO_2$ thickness $h_2$, $Ta_2O_5$ thickness $h_3$, tilt angle $\alpha$, and the unetched $Ta_2O_5$ thin film thickness $t$, all influence the diffraction efficiency and the polarization. We design the grating period $\Lambda$ and fill factor $f$ along orthogonal directions in the $xy$ plane to be equal, denoted as $\Lambda$ ($\Lambda_x = \Lambda_y$) and $f$ ($f_x = f_y$). This design choice $\Lambda_x = \Lambda_y$ ensures a spatially isotropic GMOT. $\Lambda$ and the thickness of the silver film ($h_1$) are not included in the optimization process and are fixed at 1080 nm and 200 nm, respectively.

We employ a Genetic Algorithm (GA), with a cost function as explained in Supplement 1, to optimize the parameters $f$, $h_2$, $h_3$, first assuming straight sidewalls ($\alpha = 0$) and fully etched $Ta_2O_5$ ($t = 0$ nm). When the minimum cost function is achieved, $\eta_{(1,0)}$ = 0.24 and $\xi_{SP}$ = 0.994 are obtained for optimized parameters $f$ = 0.48, $h_2$ = 520 nm, $h_3$ = 238 nm, assuming $\alpha = 0°$, $t = 0$ nm. To the best of our knowledge, this represents the highest efficiency reported to date for 2D gratings that meet the optical requirements for GMOTs. Ref. [18] numerically proposed a 2D grating with a first-order diffraction efficiency of <20%, and ref. [12] experimentally measured a first-order diffraction efficiency of approx. 21%. Using the equation for $\eta_B$, a high radiation balance of $\eta_B$ = 0.966 is obtained. The designed 2D grating enhances the optical intensity balance along the vertical axis of GMOTs with 2D gratings. It can potentially reduce the temperature of the molasses without the need for an ND filter to compensate for the intensity imbalance along the vertical axis due to low efficiency, if placed inside the chamber and illuminated by a flat-top beam. In previous studies, GMOTs utilizing low-efficiency 2D gratings required an ND filter to achieve the optical intensity balance [17,18].

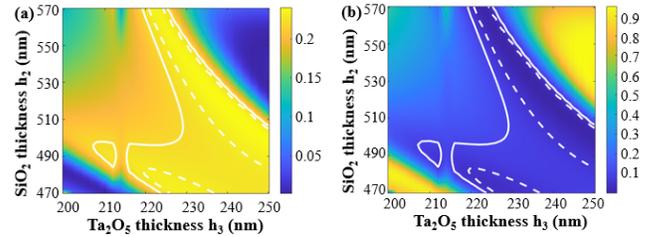

Fig. 2. Diffraction efficiency of: (a) each first order, (b) zeroth order of the grating as a function of $SiO_2$ thickness $h_2$ and $Ta_2O_5$ thickness $h_3$, with $f = 0.48$, and $\alpha = 0°$.

Fig. 2 illustrates the diffraction efficiencies of the first- and the zeroth-order beams as a function of $SiO_2$ and $Ta_2O_5$ thicknesses, with a fill factor of 0.48 and straight sidewalls. The 2D grating under circularly polarized illumination achieves a first-order diffraction efficiency of greater than 0.22, while the zeroth-order is lower than 0.10, shown in the region enclosed by the solid contour. For a smaller $SiO_2$ thickness variation in the range of 504 nm – 528 nm and $Ta_2O_5$ thickness in the range of 232 nm – 242 nm, which are readily achievable with thin film deposition techniques, the dashed contours show regions of high diffraction efficiencies above 0.23, where optimal performance occurs. Our optimized 2D grating can diffract the incident beam into four first orders, with $\xi_{SP}$ = 0.994, resulting in a high proportion of circularly polarized light with the correct handedness, $I_+$ = 0.997, as calculated from eqn. 2. This high degree of circular polarization is more than adequate to create an efficient MOT. It exceeds the values reported in previous studies of 2D gratings for GMOTs [12,18], and exhibits a high fabrication tolerance as investigated in Supplement 1.

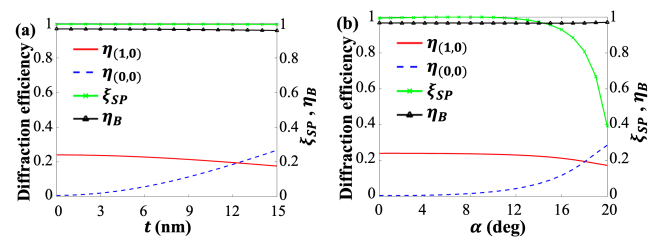

Fig. 3. Diffraction efficiency of the optimized grating, $\xi_{SP}$, and radiation balance $\eta_B$ as a function of (a) unetched $Ta_2O_5$ thin film thickness $t$ and (b) sidewall tilt angle $\alpha$, under normal incidence of circularly polarized light.

The proposed 2D grating chip can be fabricated using either electron beam lithography (EBL) or scanning beam interference lithography (SBIL), followed by reactive ion etching (RIE) [30]. Due to inevitable fabrication variations, including grating width (hole diameter) variation from lithography that results in variation in the fill factor $f$, tapered sidewalls from RIE that results in a tilt angle $\alpha$, and grating holes' height variation that results from unetched

Ta$_2$O$_5$ thin film, it is desirable that the grating performance metrics, such as the diffraction efficiency and fidelity in maintaining the correct polarization handedness upon diffraction, are tolerant to these fabrication variations.

Fig. 3(a) shows the effect of unetched Ta$_2$O$_5$ thin-film thickness on the diffraction efficiency of the first and zeroth orders, $\xi_{SP}$, and radiation balance $\eta_B$, while the other geometrical parameters are held at their optimal values ($f$ = 0.48, $h_2$ = 520 nm, $h_3$ = 238 nm), assuming $\alpha$=0°. For $t$ < 5 nm, the diffraction efficiencies of the first orders exceed 0.23, with $\xi_{SP}$ > 0.993, resulting in $\eta_B$ > 0.965. Therefore, our grating design shows an acceptable tolerance to the thickness variation of the unetched Ta$_2$O$_5$ thin film. In addition, the sidewall tilt angle reflects the fact that, during dry etching, the nanoholes deviate from a cylindrical shape to take on a frustum shape (Fig. 1(b)). Fig. 3(b) illustrates the dependence of the diffraction efficiencies, $\xi_{SP}$ and $\eta_B$, on the grating sidewall tilt angle $\alpha$, while the other geometrical parameters are held at their optimal values ($f$ = 0.48, $h_2$ = 520 nm, $h_3$ = 238 nm, $t$ = 0 nm). For $\alpha$ < 15°, the first-order diffraction efficiency is higher than 0.22, with $\xi_{SP}$ > 0.955, leading to $\eta_B$ > 0.965.

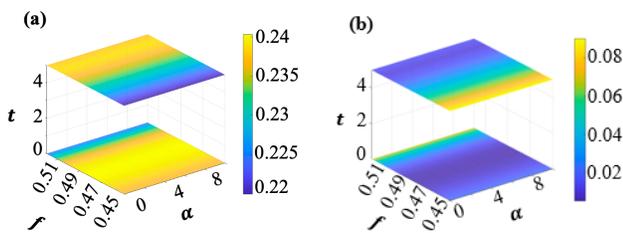

Fig. 4. Diffraction efficiency of (a) each first order, (b) zeroth order of the grating as a function of fill factor $f$ and sidewall tilt angle $\alpha$, at $t$ = 0 and $t$ = 5 $nm$, with $h_2 = 520$ $nm$, $h_3 = 238$ $nm$.

Fig. 4 illustrates the diffraction efficiencies of the first and zeroth orders as a function of the fill factor $f$ and sidewall tilt angle $\alpha$ when the unetched Ta$_2$O$_5$ film thickness is $t$ = 0 nm and $t$ = 5 nm. For $t$ = 0 nm, a high first-order diffraction efficiency exceeding 0.24 (Fig. 4(a)), and a zeroth-order efficiency less than 0.006 (Fig. 4(b)), is achieved with a fill factor of 0.473 – 0.489 and a sidewall tilt angle of 0° – 10°, demonstrating that the optimized 2D grating exhibits a high tolerance to variations in both the fill factor and sidewall tilt angle. For $t$ = 5 nm, the first-order diffraction efficiency at the fill factor $f$ = 0.48 decreases from 0.240 of $t$ = 0 nm to 0.231. However, higher diffraction efficiencies are seen at larger fill factors, which is attributed to the unetched Ta$_2$O$_5$ film increasing the effective index of the grating structure. Consequently, a higher fill factor compensates for this increase in the effective index, leading to a higher diffraction efficiency. We also investigated the use of titanium dioxide (TiO$_2$) as the grating material, paired with either Zerodur or fused silica substrates. The high-efficiency gratings showed either limited fabrication tolerances, for example $\alpha$ < 2.6°, or low degree of circular polarization, for example $\xi_{SP}$ = 0.67 (see Supplement 1).

**Acknowledgments.** This work is supported by the National Science Foundation (CAREER ECCS-2209871 and ExpandQISE-2329027).

**Disclosures.** The authors declare no conflicts of interest.
**Data availability.** The data supporting the findings of this study are available from the corresponding authors upon reasonable request.
**Supplemental document.** See Supplement 1 for supporting content.

**References**

1. E. L. Raab, M. Prentiss, A. Cable, S. Chu, and D. E. Pritchard, Phys Rev Lett **59**, 2631 (1987).
2. S. Jammi, A. R. Ferdinand, W. Zhu, G. Spektor, Z. Newman, J. Song, O. Koksal, Z. Wang, W. Lunden, D. Sheredy, P. Patel, A. Rakholia, T. C. Briles, M. Boyd, A. Agrawal, and S. Papp, (2023).
3. C. J. Picken, R. Legaie, K. McDonnell, and J. D. Pritchard, Quantum Sci Technol **4**, (2018).
4. X. Dong, S. Jin, H. Shui, P. Peng, and X. Zhou, Chinese Physics B **30**, (2021).
5. M. Vangeleyn, P. F. Griffin, I. McGregor, E. Riis, and A. S. Arnold, Optics Express, Vol. 17, Issue 16, pp. 13601-13608 **17**, 13601 (2009).
6. M. Vangeleyn, P. F. Griffin, E. Riis, and A. S. Arnold, Opt Lett **35**, 3453 (2010).
7. W. R. McGehee, W. Zhu, D. S. Barker, D. Westly, A. Yulaev, N. Klimov, A. Agrawal, S. Eckel, V. Aksyuk, and J. J. McClelland, New J Phys **23**, 013021 (2021).
8. J. Lee, J. A. Grover, L. A. Orozco, and S. L. Rolston, Journal of the Optical Society of America B **30**, 2869 (2013).
9. S. Pollock, J. P. Cotter, A. Laliotis, F. Ramirez-Martinez, and E. A. Hinds, New J Phys **13**, (2011).
10. H. Weimer, M. Müller, I. Lesanovsky, P. Zoller, and H. P. Büchler, Nature Physics 2010 6:5 **6**, 382 (2010).
11. T. Kisters, K. Zeiske, F. Riehle, and J. Helmcke, Appl Phys B **59**, 89 (1994).
12. C. C. Nshii, M. Vangeleyn, J. P. Cotter, P. F. Griffin, E. A. Hinds, C. N. Ironside, P. See, A. G. Sinclair, E. Riis, and A. S. Arnold, Nature Nanotechnology 2013 8:5 **8**, 321 (2013).
13. E. Imhof, B. K. Stuhl, B. Kasch, B. Kroese, S. E. Olson, and M. B. Squires, Phys Rev A (Coll Park) **96**, 033636 (2017).
14. J. P. Cotter, J. P. McGilligan, P. F. Griffin, I. M. Rabey, K. Docherty, E. Riis, A. S. Arnold, and E. A. Hinds, Appl Phys B **122**, 1 (2016).
15. J. Duan, X. Liu, Y. Zhou, X. B. Xu, L. Chen, C. L. Zou, Z. Zhu, Z. Yu, N. Ru, and J. Qu, Opt Commun **513**, 128087 (2022).
16. D. S. Barker, P. K. Elgee, A. Sitaram, E. B. Norrgard, N. N. Klimov, G. K. Campbell, and S. Eckel, New J Phys **25**, 103046 (2023).
17. P. F. Griffin, E. Riis, A. S. Arnold, and J. P. McGilligan, Optics Express, Vol. 23, Issue 7, pp. 8948-8959 **23**, 8948 (2015).
18. P. F. Griffin, E. Riis, A. S. Arnold, and J. P. McGilligan, JOSA B, Vol. 33, Issue 6, pp. 1271-1277 **33**, 1271 (2016).
19. S. Deshpande, P. Huft, A. Safari, C. Fang, Z. Yu, E. Oh, M. Saffman, and M. A. Kats, CLEO 2023 (2023), paper JTh2A.6 JTh2A.6 (2023).
20. J. P. McGilligan, K. Gallacher, P. F. Griffin, D. J. Paul, A. S. Arnold, and E. Riis, Review of Scientific Instruments **93**, 91101 (2022).
21. S. Bondza, C. Lisdat, S. Kroker, and T. Leopold, Phys Rev Appl **17**, (2022).
22. R. Calviac, A. Monmayrant, P. Dubreuil, L. Mazenq, S. Charlot, A. Gauguet, B. Allard, O. Gauthier-Lafaye, R. Calviac, A. Monmayrant, P. Dubreuil, L. Mazenq, S. Charlot, A. Gauguet, B. Allard, and O. Gauthier-Lafaye, JOSAB **41**, 1533 (2024).
23. M. S. C. W. K. Lindquist, (Phys. Rev., 1992), **A. 46**.
24. C. Zhang, L. Chen, Z. Lin, J. Song, D. Wang, M. Li, O. Koksal, Z. Wang, G. Spektor, D. Carlson, H. J. Lezec, W. Zhu, S. Papp, and A. Agrawal, Light Sci Appl **13**, (2024).
25. I. Mitra, Optical Materials Express, Vol. 12, Issue 9, pp. 3563-3576 **12**, 3563 (2022).
26. T. J. Bright, J. I. Watjen, Z. M. Zhang, C. Muratore, A. A. Voevodin, D. I. Koukis, D. B. Tanner, and D. J. Arenas, J Appl Phys **114**, (2013).
27. I. H. Malitson, J Opt Soc Am **55**, 1205 (1965).
28. A. Ciesielski, L. Skowronski, M. Trzcinski, and T. Szoplik, Appl Surf Sci **421**, 349 (2017).
29. Kenneth C. Johnson, "Grating Diffraction Calculator (GD-Calc ®) [Source Code]," https://doi.org/10.24433/CO.7479617.v5.
30. C. G. Chen, P. T. Konkola, R. K. Heilmann, C. Joo, and M. L. Schattenburg, https://doi.org/10.1117/12.469431 **4936**, 126 (2002).